# Valley contrasting bulk photovoltaic effect in antiferromagnetic MnPSe₃ monolayer

Qianqian Xue, Xingchi Mu, Yan Sun, Jian Zhou[*]

*Center for Alloy Innovation and Design, State Key Laboratory for Mechanical Behavior of Materials, Xi'an Jiaotong University, Xi'an, 710049 China*

[*]Email: jianzhou@xjtu.edu.cn

## Abstract

Valleytronics that uses the inequivalent electronic states at the band extrema in semiconductors have been considered to play a vital role in the future information read/write technology. In the current work, we theoretically show that sizable valley contrasting bulk photovoltaic (BPV) effect could emerge, even when the total photocurrent is symmetrically forbidden. We illustrate our theory by using a prototypical two-dimensional antiferromagnetic semiconductor, MnPSe₃ monolayer, that is $\mathcal{PT}$-symmetric ($\mathcal{P}$ and $\mathcal{T}$ refer to spatially inversion and time-reversal operators, respectively). We show that the Néel vector well controls the magnetic point group at the $\Gamma$ point, and the BPV current direction. In addition, **k**-dependent photocurrent generally arises due to the reduction of little group constraints at the valley. This leads to hidden valley-polarized photoconductivity which could reach a magnitude of 1350 μA/V², observable experimentally. We further predict that the MnPSe₃ monolayer could be two-dimensional ferrotoroidic, again depending on its Néel vector direction, which can be determined via magnetoelectric response measurements. Our work provides an exemplary platform for paving the route to future opto-spintronic and opto-valleytronic devices in a single antiferromagnetic nanomaterial.

## I. Introduction

Manipulation and control of the electronic state at band extrema, known as band valleys, have received tremendous attention during the past decade [1-4]. Different valleys in semiconductors usually locate in the momentum space with a large separation, guaranteeing their robustness against smooth geometric deformation and low energy phonon excitations. Valleytronics that uses valley degree of freedom to constitute the binary logic states (similar as the charge and spin in electronics and spintronics, respectively), holds the potential for ultrafast and efficient information and data read/write applications [5-7]. Even though early studies on electronic valleys were mainly focusing on silicon (dated back to 1970s) [8,9], recently discovered two-dimensional (2D) lattices have significantly promoted its advances [10-13]. Currently, in order to detect the valleytronic feature [14,15], one usually uses optical absorption spectrum such as circular or linear dichroism spectroscopy [16-18], and electrical approaches such as (quantum) valley Hall effect [13,19-21]. Note that electric signal is more realistic and facile for nanoelectronic devices, yet the Hall effect measurement requires depositing electrodes onto samples, which may introduce unwanted impurities or disorders.

In this work, we propose another valley contrasting feature that is stimulated by noncontacting optical illumination and can be measured and probed electrically. We discuss such optoelectronic responses via bulk photovoltaic (BPV) effect [22] in 2D antiferromagnetic (AFM) honeycomb materials [23]. The AFM systems that composes compensated spin polarization are found to be advantageous due to the absence of stray field and ultrafast spin dynamics [24]. Hence, they give rise to large information storage density and high switching efficiency in real operations. We apply group theory analysis and conduct first-principles density functional theory (DFT) calculations to show that valley contrasting BPV photocurrent could emerge in a prototypical AFM MnPSe$_3$ monolayer. The MnPSe$_3$ belongs to 2D transition metal phosphorus chalcogenides family, usually denoted as $TM$P$X_3$ ($TM$ = Cr, Mn, Fe, Co, Ni, and $X$ = S and Se). Depending on the transition metal species, this series of materials exhibit different magnetic patterns. Among them, the MnPSe$_3$ monolayer shows Néel-type AFM configuration [25], which is $\mathcal{PT}$-symmetric ($\mathcal{P}$ refers to inversion symmetry and $\mathcal{T}$ denotes time-reversal symmetry). As both $\mathcal{P}$ and $\mathcal{T}$ are broken, it holds non-degenerate valley polarizations [26]. Previous nonlinear optics theory has demonstrated that the BPV current under linearly polarized light (LPL) irradiation shows magnetic injection current (MIC) feature [27]. Our calculation suggests a sizable MIC density (1D current density on the order of 0.1–1 A/cm) could emerge under an intermediate light intensity (electric field component on the order of 0.1 MV/cm). In addition, we show that the AFM Néel vector $\mathbf{L}$ (= $\mathbf{M}_{Mn1}$ − $\mathbf{M}_{Mn2}$, Mn1 and Mn2 denote the two Mn sites in the unit cell) could effectively tune the symmetry constraints for the MIC generation. At a generic $\mathbf{k}$ point in the momentum space, the symmetry reduction could lead to hidden photocurrent generation, as revealed by a simple group theory approach. This suggests that ubiquitous valley contrasting MIC exists in the AFM MnPSe$_3$ monolayer. The Néel vector is strongly coupled with in-plane mechanical deformation, i.e., the in-plane magnetocrystalline anisotropy energy (MAE) can be manipulated via small uniaxial strains. Finally, we suggest that the MnPSe$_3$ monolayer also hosts a $\mathbf{L}$-dependent toroidal moment that can be measured by magnetoelectric responses, showing magnetically harnessed ferrotoroidicity.

## II. Computational Details

We perform DFT calculations in the Vienna *ab initio* simulation package (VASP) [28,29] that uses the generalized gradient approximation (GGA) method in the solid state Perdew-Burke-Enzerhof (PBEsol) [30] form to treat the exchange-correlation interaction. Projector augmented-wave (PAW) [31] method is used to describe the core electrons, while the valence electrons are expanded by a planewave basis set with its kinetic cutoff energy setting to be 400 eV. The first Brillouin zone (BZ) is represented by (12×12×1) Monkhorst-Pack $\mathbf{k}$-mesh grids [32], and the strong correlation on the Mn-$d$ orbital is treated by adding a Hubbard $U$ correction [33,34] with effective value of 5 eV. This has been widely adopted in previous works [25], and we note that the exact $U$ value does not affect our main conclusion. If not indicated explicitly, spin-orbit coupling (SOC) is included self-consistently in all our calculations. In order to simulate 2D materials in periodic boundary condition, we add a vacuum space of over 20 Å in the out-of-plane $z$ direction, which could eliminate the nearest neighbor image layer interactions. The convergence criteria of total energy and Hellman-Feynman force components are set as $1 \times 10^{-7}$ eV and $1 \times 10^{-3}$ eV/Å, respectively. We fit the DFT calculated electronic states by using maximally localized Wannier functions based on Mn-$s$ and $d$, P-$p$, and Se-$p$ orbitals, as implemented in the Wannier90 code [35,36], which are used to evaluate the BPV photoconductivity and magnetoelectric coupling components.

## III. Results

*Geometric, electronic structure and symmetry arguments.* The atomic structure of MnPSe$_3$ monolayer is plotted in Fig. 1(a). Geometrically, each P dimer is vertically sandwiched by six Se atoms. These P$_2$Se$_6$ moieties are embedded in the hollow sites of Mn honeycomb sublattice framework. Without considering spin polarization, it belongs to $P\bar{3}1m$ layer group ($\bar{3}m$ point group), which contains $\mathcal{C}_{3z}$ rotation, $\mathcal{C}_{2y}$ rotation, and a mirror reflection $\mathcal{M}_y$. Hence, the inversion symmetry $\mathcal{P}$ is also preserved. The Néel-type AFM configuration guarantees the unit cell being hexagonal lattice [the black rhombus in Fig. 1(a)], containing two Mn sites that carry antiparallel spin polarization. Before including SOC, the electronic states in the two spin channels (majority and minority) are degenerate [Fig. 1(b)], with the valence and conduction band valleys locating at the corner of the first BZ ($K$ and $K'$ points). The direct bandgap value is calculated to be 1.697 eV, consistent with previous reports [23].

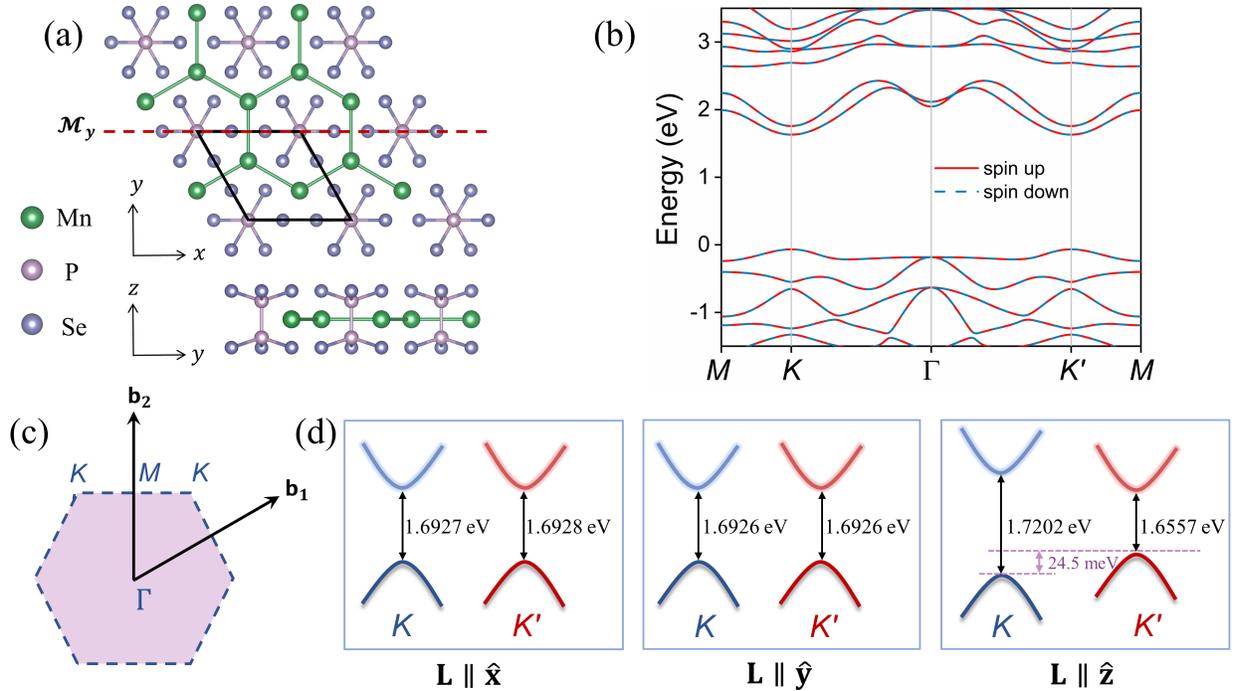

FIG. 1. (a) Top and side views of the MnPSe$_3$ monolayer. The crystalline mirror reflection $\mathcal{M}_y$ is indicated by the horizontal dashed line, and the black rhombus represents the unit cell. (b) Band dispersion along the high symmetric **k**-path without including SOC effect. (c) The first BZ, with high symmetric points of $\Gamma = (0, 0, 0)$, $K = (1/3, 1/3, 0)$, $M = (0, 1/2, 0)$, and $K' = (-1/3, 2/3, 0)$ in direct coordinates. (d) Schematic plots of band edge positions with bandgap values when the Néel vector **L** is along the $x$, $y$, and $z$ axes with SOC included. Note that each band is actually doubly degenerate due to antiunitary $\mathcal{PT}$-symmetry.

The inclusion of SOC breaks the spin rotational symmetry. In this case, the system becomes $\mathcal{PT}$-preserved, so that each band is still doubly degenerated due to its antiunitary symmetry. Since the spin angular momentum transforms as a pseudovector, its direction would determine the magnetic point group and the valley splitting. We list the basic symmetric arguments for $\mathbf{L} \parallel \hat{\mathbf{x}}$, $\mathbf{L} \parallel \hat{\mathbf{y}}$, and $\mathbf{L} \parallel \hat{\mathbf{z}}$ in Table I, ($\hat{\cdot}$ denotes the Cartesian unit vector). Here, $x$ and $y$ refer to the zigzag and armchair directions of the Mn honeycomb sublattice, respectively. Our MAE calculations reveal that in-plane spin polarization ($\mathbf{L} \parallel \hat{\mathbf{x}}$ and $\mathbf{L} \parallel \hat{\mathbf{y}}$) is energetically more favorable than the out-of-plane spin polarization ($\mathbf{L} \parallel \hat{\mathbf{z}}$) by ~0.5 meV per unit cell (or ~0.023 μJ/cm$^2$), also tabulated in Table I. Note that the $K$ and $K'$ valleys are connected via $\mathcal{T}K = K'$ or $\mathcal{M}_x K = K'$. Hence, one can deduce that when $\mathbf{L} \parallel \hat{\mathbf{y}}$, the two valleys are degenerate. On the other hand, the valley degeneracy lifts for $\mathbf{L} \parallel \hat{\mathbf{x}}$ (marginal) and $\mathbf{L} \parallel \hat{\mathbf{z}}$ (bandgap differ by 64 meV at the two valleys). Our calculations can be seen in Fig. 1(d) for schematic plots and Fig. S1 in Supplemental Material (SM) for details.

Table I. Magnetic point group, mirror reflection, and relative energies for $\mathbf{L}$ along three Cartesian axes.

| Neel vector | Magnetic point group at $\Gamma$ | Mirror reflection | Relative energy (μeV per unit cell) | Magnetic group at $K$ valley |
|---|---|---|---|---|
| $\mathbf{L} \parallel \hat{\mathbf{x}}$ | $2'/m$ | $\mathcal{M}_y$ | 0 | $m$ |
| $\mathbf{L} \parallel \hat{\mathbf{y}}$ | $2/m'$ | $\mathcal{M}_y\mathcal{T}$ | 1.63 | $m'$ |
| $\mathbf{L} \parallel \hat{\mathbf{z}}$ | $\bar{3}'m$ | $\mathcal{M}_y$ | 494 | $3m'$ |

*Valley contrasting bulk photovoltaic effect.* Next, we show that the mirror reflection difference with respect to Néel vector $\mathbf{L}$ could lead to contrasting BPV photocurrent generation. We will focus on LPL irradiation, which, according to nonlinear optics theory, generates MIC for $\mathcal{PT}$-symmetric systems [37,38]. The current density is evaluated via

$$\mathcal{J}^j = \eta_{ii}^j(0; \omega, -\omega)E^i(\omega)E^i(-\omega), \tag{1}$$

where $i$ and $j$ refer to in-plane Cartesian axes ($x$ or $y$), and $\mathbf{E}(\omega)$ is the optical alternating electric field with the angular frequency $\omega$. The MIC in $\mathcal{PT}$-systems can be viewed as a cousin effect to the normal injection current [27,37] that appears in nonmagnetic materials ($\mathcal{T}$-preserved, $\mathcal{P}$-broken) under circularly polarized light. It arises from the velocity difference between the valence and conduction bands, which increases linearly with time and saturates at the carrier lifetime. According to band theory, its length-gauge form formula is

$$\eta_{ii}^{j}(0; \omega, -\omega) = -\frac{\tau \pi e^3}{2\hbar^2} \int \frac{d^2\mathbf{k}}{(2\pi)^2} \sum_{mn} f_{mn} \Delta_{mn}^{j} g_{mn}^{ii} \delta(\omega_{mn} - \omega). \quad (2)$$

Here, $f_{mn} = f_m - f_n$ and $\Delta_{mn}^{j} = v_{mm}^{j} - v_{nn}^{j}$ measure the Fermi-Dirac occupation and group velocity differences between the band $m$ and $n$, respectively. The Kronecker delta function $\delta(\omega_{mn} - \omega)$, represented by the Lorentz function with a broadening factor of 0.02 eV, guarantees the energy conservation law, with $\hbar\omega_{mn} = \hbar\omega_m - \hbar\omega_n$ referring to the eigenenergy difference. The MIC generation is scaled by quantum metric tensor $g_{mn}^{ii} = 2\sum_{\mu,\nu} \text{Re}\left(r_{m_\mu n_\nu}^{i} r_{n_\nu m_\mu}^{i}\right)$, where $\mu$ and $\nu$ represent the degenerate band indices, and the position matrix is $r_{nm}^{i} = \langle n|\hat{r}^i|m\rangle = \frac{\langle n|\hat{v}^i|m\rangle}{i\omega_{nm}}$. All these quantities are $\mathbf{k}$-dependent which are omitted for clarity reason. According to previous works [38], we take the carrier lifetime $\tau$ to be a universal value of 0.2 ps, comparable to most experimental observations in 2D materials. This carrier lifetime can be effectively controlled by the sample quality, disorder level, temperature, etc., and can be characterized by the electrical conductance according to the Drude model. The integral is performed in the whole 2D first BZ. We take the effective thickness as $d = 0.6$ nm, measured from its bulk counterpart. Then, we divide the 2D photoconductivity [$\mu$A·nm/V$^2$, according to Eq. (2)] by $d$, so that it adopts the conventional 3D photoconductivity unit ($\mu$A/V$^2$).

Before performing DFT calculations, we briefly analyze the magnetic point group for each case and their implication for BPV photocurrents. The highest symmetry exists when the Néel vector is parallel to $z$, $\mathbf{L} \parallel \hat{\mathbf{z}}$, and the system belongs to magnetic point group $\overline{3}'m = C_{3v}\otimes\mathcal{PT}$. Since we are focus on the MIC (injection current under LPL) that is invariant under $\mathcal{PT}$, we focus on the $C_{3v}$ point group (character table can be found in Tables S1). For the electric field and current in the 2D ($xy$) plane, the irreducible representation for current and second order symmetric field are $\Gamma_{\mathcal{J}} = E$ and $\Gamma_{(\mathbf{EE})^s} = A_1 \oplus E$. Hence, one has $\Gamma_{\mathcal{J}}\otimes\Gamma_{(\mathbf{EE})^s} = A_1 \oplus A_2 \oplus 2E$, allowing only one nonzero independent MIC component, which will be shown to be $\eta_{xx}^x = -\eta_{yy}^x = -\eta_{xy}^y$ and $\eta_{xx}^y = \eta_{yy}^y = \eta_{xy}^x = 0$. If we focus on the valley $K$ (or $K'$), the spatial inversion is no longer preserved, giving its magnetic little group of $3m' = C_3 \oplus C_s \mathcal{T}$. Thus, the symmetry argument at each valley for MIC generation follows $C_3$ (since MIC is $\mathcal{PT}$-symmetric which is inconsistent with $C_s \mathcal{T}$), yielding $\Gamma_{\mathcal{J}} \otimes \Gamma_{(\mathbf{EE})^s} = E \otimes (A \oplus E) = 2A \oplus 2E$ with two allowed and independent MIC components (Table S2). This indicates that momentum-dependent hidden MIC exists. This is akin to the hidden spin polarization (or spin Hall effect) as discovered locally in centrosymmetric ionic compounds and antiferroelectric materials [39-43], which arises in the real space due to the reduced symmetry constraints on each sector. Similar confinements can be found for in-plane Néel vector, which breaks the three-fold rotation $C_{3z}$. The $\mathbf{L} \parallel \hat{\mathbf{x}}$ belongs to $2'/m = C_s \otimes \mathcal{PT}$, and the allowed MIC generation can be deduced from the $C_s$ point group (see its character table in Table S3). We then have $x$-flowing MIC satisfying $\Gamma_{\mathcal{J}} \otimes \Gamma_{(\mathbf{EE}^*)^s} = A' \otimes (2A' \oplus A'') = 2A' \oplus A''$. The two independent MIC would be $\eta_{xx}^x$ and $\eta_{yy}^x$. The $\mathbf{L} \parallel \hat{\mathbf{y}}$ is $2/m' = C_2 \otimes \mathcal{PT}$, and one can perform similar arguments to yield same results as in $\mathbf{L} \parallel \hat{\mathbf{y}}$. At the valleys, their $180°$-rotation symmetry ($2'$ or $2$) is no longer preserved. Hence, they both exhibit valley-dependent finite MIC components that are forbidden in the whole system.

The switching of $\mathbf{L}$ strongly affects the velocity texture distribution in $\mathbf{k}$-space, so that the MIC direction depends on the Néel vector [see Eq. (2)]. In detail, when $\mathbf{L} \parallel \hat{\mathbf{x}}$ and $\mathbf{L} \parallel \hat{\mathbf{z}}$, the $\mathcal{M}_y$ reflection assigns $\mathcal{M}_y \Delta^y(k_x, k_y) = -\Delta^y(k_x, -k_y)$. On the contrary, for the $\mathbf{L} \parallel \hat{\mathbf{y}}$ case, we have $\mathcal{M}_y \mathcal{T} \Delta^x(k_x, k_y) = -\Delta^x(-k_x, k_y)$. Since the quantum metric tensor is almost unchanged in these cases, one easily deduces that the $x$-flowing MIC is forbidden when $\mathbf{L} \parallel \hat{\mathbf{y}}$, while the $y$-flowing MIC is zero when $\mathbf{L} \parallel \hat{\mathbf{x}}$ or $\mathbf{L} \parallel \hat{\mathbf{z}}$. Note that here we assume that the $x$- (or $y$-) polarized LPL is illuminated. For a general polarization angle, such symmetry arguments may be changed, and the final MIC will be a linear combination from the $x$-LPL and $y$-LPL.

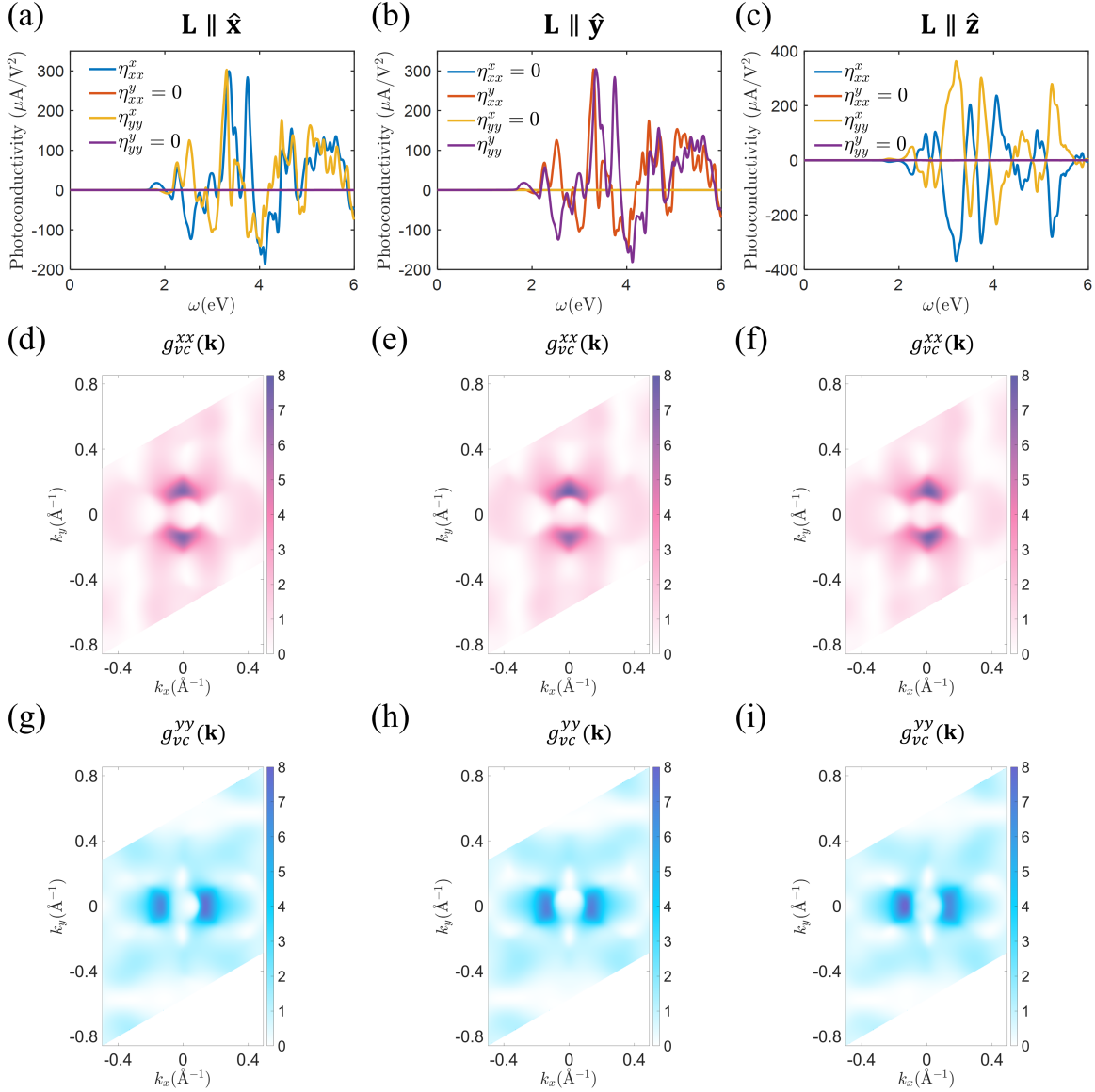

FIG. 2. Calculated MIC photoconductivity under $x$ and $y$-polarized LPL for (a) $\mathbf{L} \parallel \hat{\mathbf{x}}$, (b) $\mathbf{L} \parallel \hat{\mathbf{y}}$, and (c) $\mathbf{L} \parallel \hat{\mathbf{z}}$. In (c), we note that $\eta_{xx}^x(\omega) = -\eta_{yy}^x(\omega)$, arising from the $\mathcal{C}_{3z}$ rotation. Such symmetry is broken when $\mathbf{L}$ lies in-plane. Quantum metric distribution between the top valence and bottom conduction bands $g_{vc}^{xx}(\mathbf{k})$ over the first BZ for (d) $\mathbf{L} \parallel \hat{\mathbf{x}}$, (e) $\mathbf{L} \parallel \hat{\mathbf{y}}$, and (f) $\mathbf{L} \parallel \hat{\mathbf{z}}$, and $g_{vc}^{yy}(\mathbf{k})$ for (g) $\mathbf{L} \parallel \hat{\mathbf{x}}$, (h) $\mathbf{L} \parallel \hat{\mathbf{y}}$, and (i) $\mathbf{L} \parallel \hat{\mathbf{z}}$.

Our first-principles calculations confirm the above qualitative results. As shown in Figs. 2(a)–2(c), we plot the calculated MIC generation photoconductivity. One clearly observes that $\eta_{ii}^y = 0$ for $\mathbf{L} \parallel \hat{\mathbf{x}}$ and $\mathbf{L} \parallel \hat{\mathbf{z}}$ ($i = x$ or $y$). When $\mathbf{L}$ is switched to along $y$, the $\eta_{ii}^x$ becomes zero. For the symmetrically allowed current, the magnitude of photoconductivity reaches ~360 μA/V² ($\mathbf{L} \parallel \hat{\mathbf{z}}$).

It indicates that if we take the electric field magnitude of 0.1 V/nm (at the photon energy of 3.2 eV, or wavelength of 387 nm), corresponding to $1.33 \times 10^9$ W/cm$^2$ light intensity, one could generate ~3.6 μA/nm$^2$ current density. Across the lateral size of 1 nm (note that the effective thickness is $d = 6$ Å), the current reaches 2.16 μA. Normal to the MIC, no *net* photocurrent can be detected. This vividly suggests that switching magnetic moment direction could drastically rotate the MIC generation direction. Such a large contrast can be directly measured via closed-circuit current or open-circuit voltage. The quantum metric between the top valence band and bottom conduction band (both doubly degenerate) $g_{vc}^{xx}(\mathbf{k})$ and $g_{vc}^{yy}(\mathbf{k})$ for $\mathbf{L} \parallel \hat{\mathbf{x}}, \mathbf{L} \parallel \hat{\mathbf{y}}, \mathbf{L} \parallel \hat{\mathbf{z}}$ are shown in Figs. 2(d)–2(i). We can see symmetry argument of $g_{vc}^{ii}(k_x, k_y) = g_{vc}^{ii}(k_x, -k_y)$, ($i = x$ or $y$) for $\mathcal{M}_y$ and $g_{vc}^{ii}(k_x, k_y) = g_{vc}^{ii}(-k_x, k_y)$ for $\mathcal{M}_y \mathcal{T}$. In addition, we note that the time-reversal symmetry $\mathcal{T}$ that flips the Néel vector $\mathbf{L}$ (e.g., between $\mathbf{L} \parallel \hat{\mathbf{x}}$ and $\mathbf{L} \parallel -\hat{\mathbf{x}}$, or from $\mathbf{L} \parallel \hat{\mathbf{z}}$ to $\mathbf{L} \parallel -\hat{\mathbf{z}}$) also reverses the MIC generation while keeping the magnitude, as the $\eta_{ii}^j$ is scaled by velocity operator and being $\mathcal{T}$-odd (see Fig. S2).

We then show that sizable valley contrasting MIC emerges even when the net current is zero. In order to explicitly see this, we plot the **k**-resolved MIC contributions, namely, the integrand of Eq. (2), for the symmetrically forbidden currents [Figs. 3(a)–3(c)]. The incident photon frequency is selected to be $\hbar\omega = 1.8$ eV, slightly above the bandgap. In all these cases, the mirror symmetry constraints (Table I) are clearly observed. Main contributions appear in the vicinity of $K$ (and $K'$) valleys, in agreement with the joint density of states at this incident energy. Even though the contributions around one valley show both positive and negative ridges, their summation is nonzero. In Figs. 3(d)–(f), we plot the valley-dependent MIC, which is integrated in the triangular **k**-space around each valley, whose area equals to half of BZ. We see that both valleys contribute significant MIC generation, reaching a photoconductivity of ~1500 μA/V$^2$ (for $\mathbf{L} \parallel \hat{\mathbf{x}}$ and $\mathbf{L} \parallel \hat{\mathbf{z}}$) or ~600 μA/V$^2$ (for $\mathbf{L} \parallel \hat{\mathbf{y}}$). In each case, the MICs from two valleys flow oppositely with the same magnitude, giving zero net MIC generation. This hidden valley-dependent BPV current has been largely overlooked previously, and may find its potential applications in 2D valleytronic devices. We propose that this result provides another scheme to use the valley current, in addition to the electrically triggered (quantum) valley Hall effect. Experimentally, one could design a magnetic heterostructure [Fig. 3(g)] to separate and measure such valley-dependent MIC. Valley contrasting MIC could accumulate at a domain wall between two AFM configurations that are time-reversal with each other (e.g., $\mathbf{L} \parallel \hat{\mathbf{z}}$ and $\mathbf{L} \parallel -\hat{\mathbf{z}}$), so that the current contributed from a specific valley $K$ (or $K'$) in both domains flow to their boundary. Similar valley-dependent MIC also exists for the symmetrically allowed components, as plotted in Fig. S3. The two valleys also hold oppositely flowing MIC, but they do not completely cancel each other.

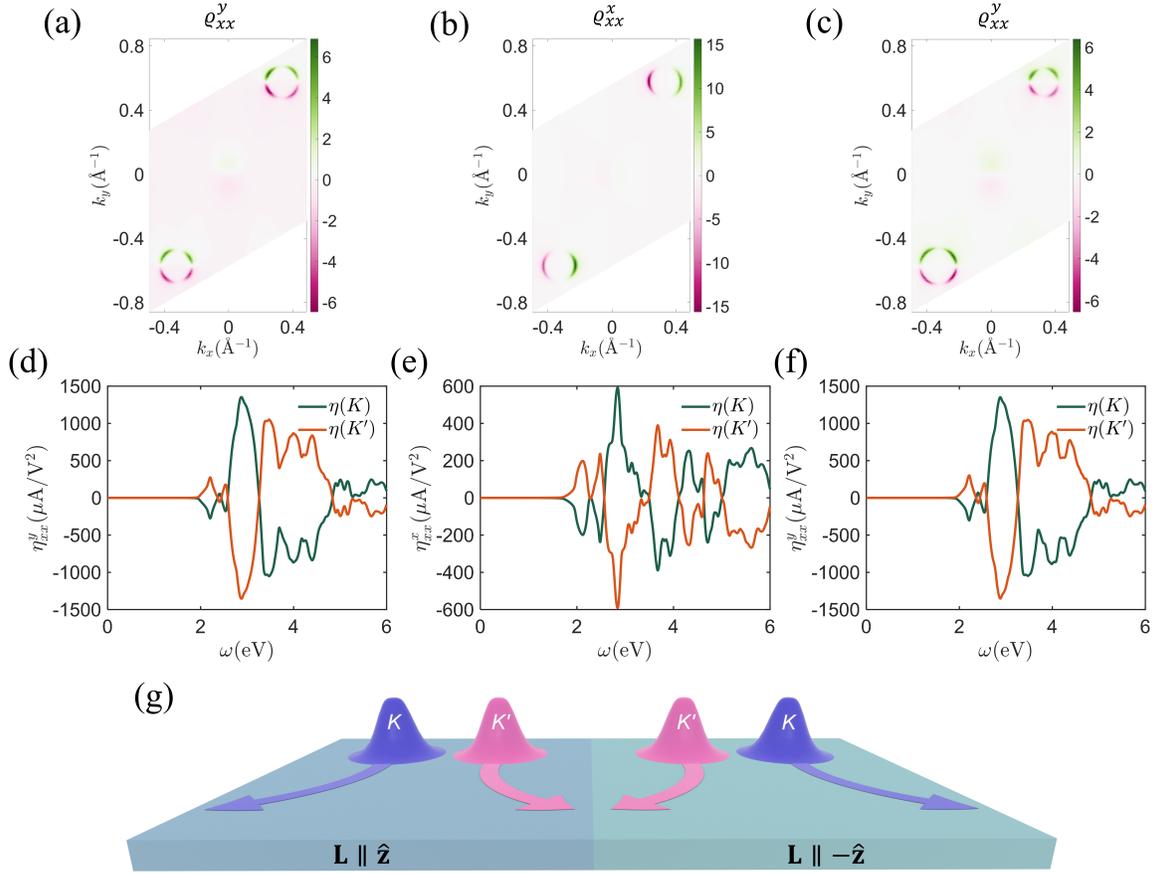

FIG. 3. BZ contribution of MIC integrand $\varrho_{xx}^{j}(\mathbf{k}, \omega) = \sum_{mn} f_{mn} \Delta_{mn}^{j} g_{mn}^{xx} \delta(\omega_{mn} - \omega)$ under $x$-PLP irradiation for (a) $\mathbf{L} \parallel \hat{\mathbf{x}}$ ($j = y$), (b) $\mathbf{L} \parallel \hat{\mathbf{y}}$ ($j = x$), and (c) $\mathbf{L} \parallel \hat{\mathbf{z}}$ ($j = y$). Here the incident photon energy is $\hbar\omega = 1.8$ eV. (d)–(f) plot their corresponding valley contrasting MIC contributions. (g) Schematic plot of magnetic heterostructure with two AFM configurations that are time-reversal with each other (e.g., between $\mathbf{L} \parallel \hat{\mathbf{z}}$ and $\mathbf{L} \parallel -\hat{\mathbf{z}}$). Valley-dependent BPV current would accumulate at their domain boundary.

*MAE modulation under strains.* One may wonder how to harness the in-plane MAE ($E_{\text{MAE}} = E_{\mathbf{L}\parallel\hat{\mathbf{x}}} - E_{\mathbf{L}\parallel\hat{\mathbf{y}}}$), so that the Néel vector $\mathbf{L}$ can be pinned along $x$ or $y$. We show that a uniaxial strain could further split the energy difference between $\mathbf{L} \parallel \hat{\mathbf{x}}$ and $\mathbf{L} \parallel \hat{\mathbf{y}}$. Our numerical results are shown in Fig. 4. At the equilibrium (strain-free) state, the $\mathbf{L} \parallel \hat{\mathbf{x}}$ is almost degenerate with $\mathbf{L} \parallel \hat{\mathbf{y}}$ ($E_{\text{MAE}} =$

−1.63 μeV per unit cell). Under tensile strain along $x$ ($\varepsilon_{xx}$), the $E_{MAE}$ is negative, so that the Néel vector **L** prefers the $x$ direction. On the other hand, the $y$-tensile strain increases the $E_{MAE}$, aligning the **L** along $y$. In both cases, a small strain of 3% (about elastic energy of 49 meV in one unit cell) could enhance the MAE magnitude to be ~35 μeV per unit cell, which is large enough to be distinguished in low temperature experiments. Furthermore, we find that such a small strain will not significantly change the band structure in these cases, and the MIC photoconductivity marginally change their values.

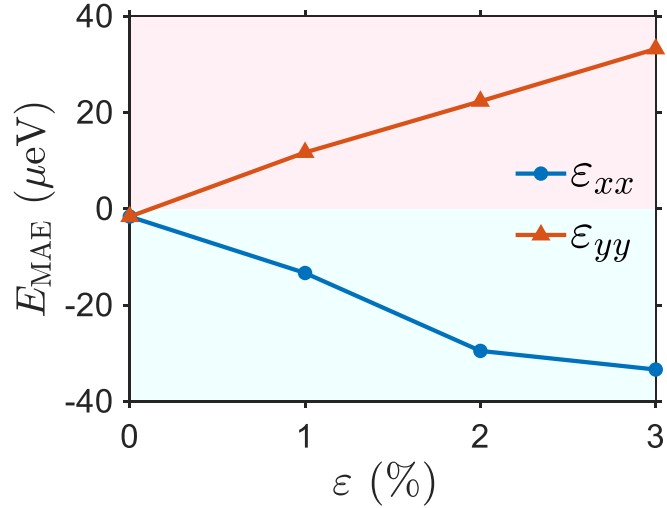

FIG. 4. Total energy difference (per unit cell) between **L** ∥ $\hat{\mathbf{x}}$ and **L** ∥ $\hat{\mathbf{y}}$ under uniaxial strain $\varepsilon$ along $x$ and $y$. The out-of-plane AFM configuration **L** ∥ $\hat{\mathbf{z}}$ is always much higher in energy under such small strains.

***L-dependent ferrotoroidicity and magnetoelectric responses.*** In addition to the optically induced nonlinear current, we now show that the ferrotoroidicity is also sensitive to the Néel vector direction. Ferrotoroidicity has been discovered to be another primary ferroic order, compensating the ferroelasticity ($\mathcal{P}$-even, $\mathcal{T}$-even), ferroelectricity ($\mathcal{P}$-odd, $\mathcal{T}$-even), and ferromagnetism ($\mathcal{P}$-even, $\mathcal{T}$-odd) [44,45]. It reverses its sign under either $\mathcal{P}$ or $\mathcal{T}$, and is defined by toroidal moment $\mathbf{t} = \sum_i \mathbf{r}_i \times \mathbf{s}_i$, where $\mathbf{r}_i$ and $\mathbf{s}_i$ represent the position and spin vectors of ion-$i$, and the summation runs over all sites in the supercell. Previous theoretical and experimental works have disclosed a few ferrotoroidic bulk materials, such as $LiCoPO_4$ [46], $LiFeSi_2O_6$ [47], and defective $SrTiO_3$ [48]. Here, we suggest that the 2D $MnPSe_3$ monolayer also holds ferrotoroidic order with sizable out-of-plane toroidal moment, if the Néel vector is pointing away from $y$. As shown in Fig. 5(a), we schematically depict the Mn sublattice distributions in a rectangular supercell, which, without loss of generality, is uniaxially strained. The supercell contains four Mn sites, which locate on two co-centered circles with radius of $r_1$ and $r_2$. At the equilibrium state, these two green circles are identical, $r_1 = r_2$, and the angle $\theta = 30°$. Geometrically, if $x$-tensile strain is applied, $\theta$ reduces and $r_1 < r_2$; $y$-tension will increase $\theta$ and makes $r_1 > r_2$.

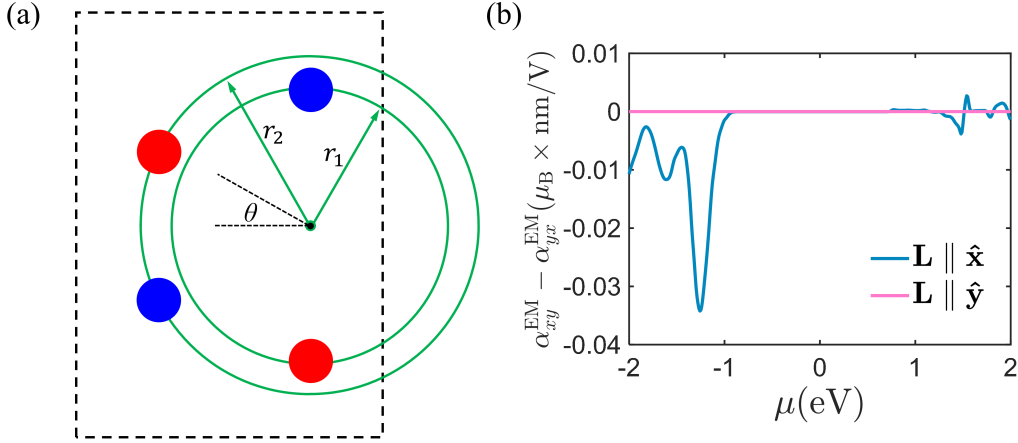

FIG. 5. (a) Schematic plot of ferrotoroidicity in a rectangular supercell. Blue and red circles represent the antiparallel spin polarized Mn sites, and the two green circles (radius $r_1$ and $r_2$) are co-centered. Detailed explanation can be found in the main text. (b) Calculated magnetoelectric coefficient for $\mathbf{L} \parallel \hat{\mathbf{x}}$ and $\mathbf{L} \parallel \hat{\mathbf{y}}$. The abscissa axis denotes the chemical potential relative to the Fermi energy.

We can directly estimate the toroidal moment $\mathbf{t}$ in this supercell setup. Note that similar as the electric polarization $\mathbf{P}$, here $\mathbf{t}$ is also multi-valued with respect to a quanta, depending on the choice of origin. Nonetheless, we can use this position-spin cross product definition to determine its existence. When the Mn spin polarization is along $x$, namely, $\mathbf{s} = (\pm s_x, 0,0)$, it can be directly deduced that $\mathbf{t} = (0,0,t_z)$ and $t_z = -\sum_{i=1}^{4} y_i s_{i,x} = -2s_x(r_1 - r_2 \sin\theta) \neq 0$. Hence, it shows a nonzero $z$ component toroidal moment. If $\mathbf{s} = (0, \pm s_y, 0)$, one could easily find $\mathbf{t} = (0,0,0)$, constrained by the $\mathcal{M}_y \mathcal{T}$ operation, even though the supercell is uniaxially strained. This can also be understood by analyzing the $C_{2h}$ point group (Table S4). The $\mathbf{L} \parallel \hat{\mathbf{x}}$ belongs to the magnetic point group $2'/m$, which gives negative character for both $C_2$ rotation and inversion $i$. Hence, one has $\Gamma_m = B_u$ in which all other elements are represented by $+1$. The vertically aligned electric field and spin polarization are presented by $\Gamma_{\mathbf{E}} = A_u$ and $\Gamma_{\mathbf{M}} = B_g$ (or $\Gamma_{\mathbf{E}} = B_u$ and $\Gamma_{\mathbf{M}} = A_g$), respectively. Then we have $\Gamma_m \otimes \Gamma_{\mathbf{E}} \otimes \Gamma_{\mathbf{M}} = A_g$, which is symmetrically allowed. On the other hand, the magnetic point group of $2/m'$ for $\mathbf{L} \parallel \hat{\mathbf{y}}$ gives $\Gamma_m = B_g$. Thus, $\Gamma_m \otimes \Gamma_{\mathbf{E}} \otimes \Gamma_{\mathbf{M}} = A_u$, being forbidden. Note that flipping Néel vector $\mathbf{L}$ corresponds to time-reversal operation, thus the ferrotoroidic vector $\mathbf{t}$ is also reversed between $\mathbf{L}$ and $-\mathbf{L}$.

The ferrotoroidicity can be reflected by the nondiagonal elements of magnetoelectric response coefficient tensor $\alpha^{\text{EM}}$, defined as $M_j = \alpha_{ij}^{\text{EM}} E_i$. According to Kubo perturbation theory, $\alpha^{\text{EM}}$ can be calculated by

$$\alpha_{ij}^{\text{EM}} = \frac{eA}{\hbar} \int \frac{d^2\mathbf{k}}{(2\pi)^2} \operatorname{Im} \sum_{n,l} \frac{f_{ln} v_{nl}^i m_{ln}^j}{(\omega_{ln} + i/\tau)^2} + \frac{eA}{\hbar} \tau \int \frac{d^2\mathbf{k}}{(2\pi)^2} \sum_n v_{nn}^i m_{nn}^j \delta(\omega_n - \mu).$$ (3)

Here, $m_{ln}^j = \langle l\mathbf{k}|\hat{m}^j|n\mathbf{k}\rangle \simeq -2\langle l\mathbf{k}|\hat{s}^j|n\mathbf{k}\rangle$ is the magnetic moment matrix element, and only spin contribution is taken account in this work. $A$ refers to the area of unit cell, hence $\alpha_{ij}^{\text{EM}}$ measures the total magnetic moments in the unit cell induced by an in-plane static electric field $E_i$ (also called as Rashba-Edelstein coefficient [49,50]). The first term arises from the interband contribution, while the second term evaluates the intrinsic contributions from the Fermi surface, being $\tau^1$-dependent (similar as the MIC generation). According to previous discussions [51], $T_k \sim \epsilon_{ijk} \alpha_{ij}^{\text{EM}}$ where $\epsilon_{ijk}$ is the Levi-Civita symbol (with Einstein summation convention). Thus, we plot the nondiagonal difference $\left( \alpha_{xy}^{\text{EM}} - \alpha_{yx}^{\text{EM}} \right)$ as a function of chemical potential $\mu$ in Fig. 5(b). It is clear that when $\mathbf{L} \parallel \hat{\mathbf{y}}$, the $\left( \alpha_{xy}^{\text{EM}} - \alpha_{yx}^{\text{EM}} \right) = 0$, consistent with previous discussions. The $\mathbf{L} \parallel \hat{\mathbf{x}}$ pattern gives finite magnetoelectric responses. When the chemical potential lies inside the bandgap, only extrinsic interband contribution exists, which is found to be 0.03 $\mu_B \times$nm/V ($\mu_B$ is Bohr magneton). Upon $n$- or $p$-type doping, the intrinsic Fermi surface term comes in, which significantly increases $\left( \alpha_{xy}^{\text{EM}} - \alpha_{yx}^{\text{EM}} \right)$ to the order of 0.01 $\mu_B \times$nm/V. Thus, an intermediate electric field with 0.1 V/nm strength yields about $10^{-3}$ $\mu_B$ magnetic moment variation, being sufficiently large for experimental observation. Considering the magnetic exchange parameter $J_{\text{ex}}$ of MnPSe$_3$ is 0.12 meV/$(\mu_B)^2$ [25], we can estimate the effective magnetic field of the magnetoelectric coupling to be $\sim$20 mT per (V/nm). This magnetoelectric responses serve as an indirect and complementary demonstration for magnetic moment direction dependent ferrotoroidicity.

## IV. Discussion

Before conclusion, we would like to remark a few points. In addition to charge current, recent advances have been extending the BPV effect into spin degree of freedom, namely, bulk spin photovoltaic (BSPV) generation [38,52,53]. Here, we follow this route and compute the BSPV photoconductivity. Previous works [27] have demonstrated that the LPL-induced spin photocurrent belongs to the shift current nature for $\mathcal{PT}$-symmetric systems, rather than MIC mechanism for the electric charge current. The spin current operator is defined as $\hat{j}^{ij} = \frac{1}{2}\left(\hat{v}^i \hat{s}^j + \hat{s}^j \hat{v}^i\right)$, where we adopt spin parallel to Néel vector, namely, $j$ is along $\mathbf{L}$. Our calculation results are plotted in Fig. S4. We find that no matter $\mathbf{L}$ is parallel to $x$, $y$, or $z$, the BSPV current always flows along $y$, making the $x$-propagating spin current symmetrically forbidden. This is because the spin current operator contains a surplus spin operator that transforms as pseudovector, compared with electric charge current operator. Nevertheless, valley-dependent spin photocurrents still exist, even though in the forbidden net spin current situation.

The SOC effect plays an essential role in not only breaking the spin rotational symmetry, but significantly affecting the MIC generation. In order to show this, we adjust the SOC strength by multiplying a tuning factor $\lambda \in [0,1]$. Here $\lambda = 0$ turns off the SOC effect, and $\lambda = 1$ refers full SOC. As shown in Fig. S5, zero MIC is generated when the SOC is absent. We find that as SOC is gradually increased, the symmetrically allowed MIC photoconductivity almost linearly enhances. Note that when SOC is turned on, the spin magnetic quantum number is not conserved and one cannot calculate the MIC from the two spin channels separately. Such SOC variation effects do not largely affect the BSPV current, which remains to be finite regardless with $\lambda$. We note that such SOC tunable BPV effect in AFM $\mathcal{PT}$-symmetric systems is different from the nonmagnetic ($\mathcal{T}$-symmetric, $\mathcal{P}$-broken) materials [38], where spin photocurrent linearly enhances with $\lambda$ but the electric charge current remains almost unchanged.

The AFM pattern also determines the symmetry constraints. In this work, we focus on the Néel pattern in the MnPSe$_3$ monolayer, as determined by recent experiments [54,55]. If other transition metals are used, e.g., FeP$X_3$ and CrP$X_3$ ($X$ = S or Se), stripe or zigzag AFM pattern could become energetically optimal [23,25,56,57]. In these cases, the system is $\mathcal{P}$-symmetric, rather than $\mathcal{PT}$. According to previous discussions, the second order nonlinear BPV current vanishes for $\mathcal{P}$-symmetric systems, regardless the local spin polarization directions. In addition, the band extrema in these cases do not locate at the $K$ (or $K'$) point, hence we cannot determine valley-polarized BPV effect here, even though **k**-resolved BPV photocurrents do not completely vanish.

Inversion symmetry could also preserve in Néel AFM patterns when we stack two monolayers together and form a MnPSe$_3$ bilayer. In order to illustrate this, we calculate the BPV photoconductivity, and the results are shown in Fig. S6. One could see that depending on the spin polarization patterns between the two monolayers, the whole system can be either $\mathcal{P}$ or $\mathcal{PT}$. Hence, zero or finite photocurrent emerges in these cases. This is akin to the recently proposed sliding ferroelectricity [58-60] that arises from atomic interfacial mismatch between the two layers, while here it is the magnetic order that is mismatched at the interface. Such interlayer spin order adjusted symmetry in bilayer AFM materials is beyond the scope in the current work and will be discussed in detail elsewhere.

## V. Conclusion

In summary, we conduct group theory and first-principles calculations on MnPSe$_3$ monolayer to show that photoinduced MIC generation in AFM $\mathcal{PT}$-symmetric materials sensitively depends on the spin polarization Néel vector **L**. The symmetry arguments, especially mirror reflection, vary by switching **L**. In addition, we show that sizable valley contrasting photocurrents could exist in AFM $\mathcal{PT}$-symmetric materials, even though the net MIC component is symmetrically constrained to be zero. The Néel vector direction can be well-tuned by applying external uniaxial stress, which also harnesses the toroidal moment and the magnetoelectric coupling. This work provides an in-depth examination on the AFM magnetic order implications on various electrical and optical responses, and paves the route to realizing nanoscale optoelectronic, optospintronic, and optovalleytronic devices with ultrafast kinetics.

**Acknowledgments.** This work was supported by the National Natural Science Foundation of China (NSFC) under Grant No. 11974270. The computational resources from HPC platform of Xi'an Jiaotong University are also acknowledged.

# Supplemental Material for Valley contrasting bulk photovoltaic effect in antiferromagnetic MnPSe₃ monolayer

Qianqian Xue[1], Xingchi Mu[1], Yan Sun[1], Jian Zhou[1,*]

[1]Center for Alloy Innovation and Design, State Key Laboratory for Mechanical Behavior of Materials, Xi'an Jiaotong University, Xi'an, 710049 China

*Email: jianzhou@xjtu.edu.cn

## 1. Supplemental Tables.

Table S1. Character table for $C_{3v}$, the symmetric ($\Gamma_{(\mathbf{EE})^s}$) and antisymmetric ($\Gamma_{(\mathbf{EE})^a}$) electric field representations are generated in 2D $xy$ plane.

|  | $E$ | $2C_3(z)$ | $3\sigma_v$ | basic function |
|---|---|---|---|---|
| $A_1$ | 1 | 1 | 1 | $z$ |
| $A_2$ | 1 | 1 | $-1$ | $R_z$ |
| $E$ | 2 | $-1$ | 0 | $(x, y)\,(R_x, R_y)$ |
| $\Gamma_{(\mathbf{EE})^s}$ | 3 | 0 | 1 |  |
| $\Gamma_{(\mathbf{EE})^a}$ | 1 | 1 | $-1$ |  |

Table S2. Character table for $C_3$  ($\omega = e^{i\frac{2\pi}{3}}$).

|  | $E$ | $C_3(z)$ | $C_3(z)^2$ | basic function |
|---|---|---|---|---|
| $A$ | 1 | 1 | 1 | $z,\ R_z$ |
| $E$ | 1 | $\omega$ | $\omega^*$ | $x + iy;\ R_x + iR_y$ |
|  | 1 | $\omega^*$ | $\omega$ | $x - iy;\ R_x - iR_y$ |
| $\Gamma_{(\mathbf{EE})^s}$ | 3 | 0 | 0 |  |
| $\Gamma_{(\mathbf{EE})^a}$ | 1 | 1 | 1 |  |

Table S3. Character table for $C_s$  (assuming mirror vertical to $y$).

|  | $E$ | $\sigma_h$ | basic function |
|---|---|---|---|
| $A'$ | 1 | 1 | $x, z, R_y$ |
| $A''$ | 1 | $-1$ | $y, R_x, R_z$ |
| $\Gamma_{(\mathbf{EE})^s}$ | 3 | 1 |  |
| $\Gamma_{(\mathbf{EE})^a}$ | 1 | $-1$ |  |

Table S4. Character table for $C_{2h}$ (assuming axis along $y$).

|  | $E$ | $C_2(y)$ | $i$ | $\sigma_h$ | basic function |
|---|---|---|---|---|---|
| $A_g$ | 1 | 1 | 1 | 1 | $R_y$ |
| $B_g$ | 1 | $-1$ | 1 | $-1$ | $R_x, R_z$ |
| $A_u$ | 1 | 1 | $-1$ | $-1$ | $y$ |
| $B_u$ | 1 | $-1$ | $-1$ | 1 | $x, z$ |

## 2. Supplemental Figures

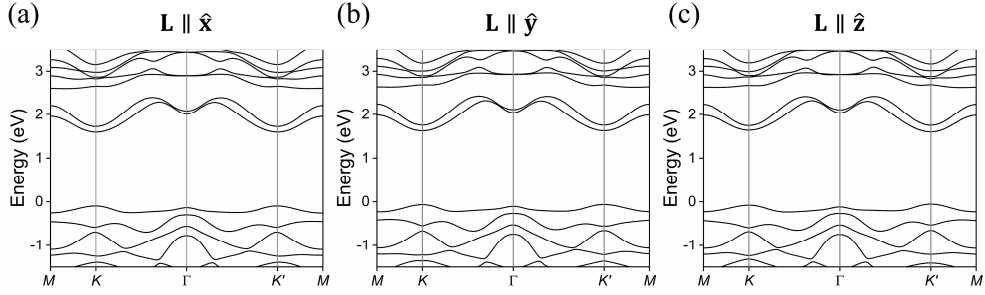

FIG. S1. Band dispersion along the high symmetric **k**-path including SOC effect. (a) $\mathbf{L} \parallel \hat{\mathbf{x}}$, (b) $\mathbf{L} \parallel \hat{\mathbf{y}}$, and (c) $\mathbf{L} \parallel \hat{\mathbf{z}}$.

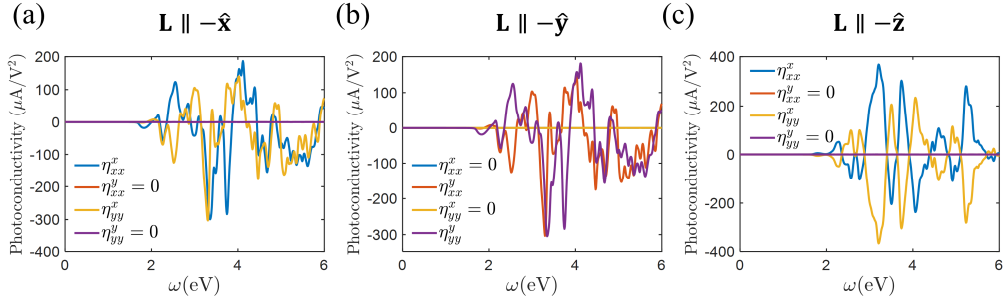

FIG. S2. MIC photoconductivity under $x$ and $y$-polarized LPL for (a) $\mathbf{L} \parallel -\hat{\mathbf{x}}$, (b) $\mathbf{L} \parallel -\hat{\mathbf{y}}$, and (c) $\mathbf{L} \parallel -\hat{\mathbf{z}}$. One sees that time-reversal operation reverses current.

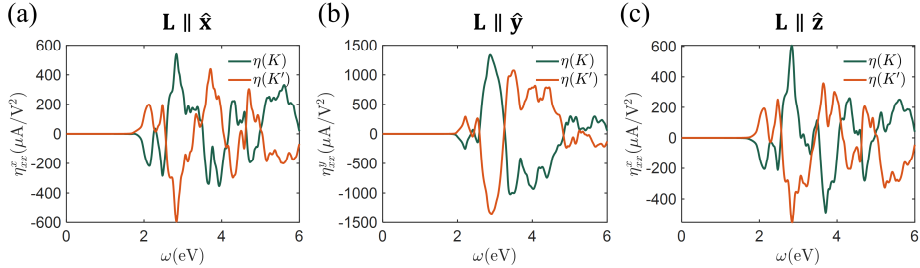

FIG. S3. The MIC photoconductivity under $x$-polarized LPL for (a) $\mathbf{L} \parallel \hat{\mathbf{x}}$ ($j = x$), (b) $\mathbf{L} \parallel \hat{\mathbf{y}}$ ($j = y$), and (c) $\mathbf{L} \parallel \hat{\mathbf{z}}$ ($j = x$).

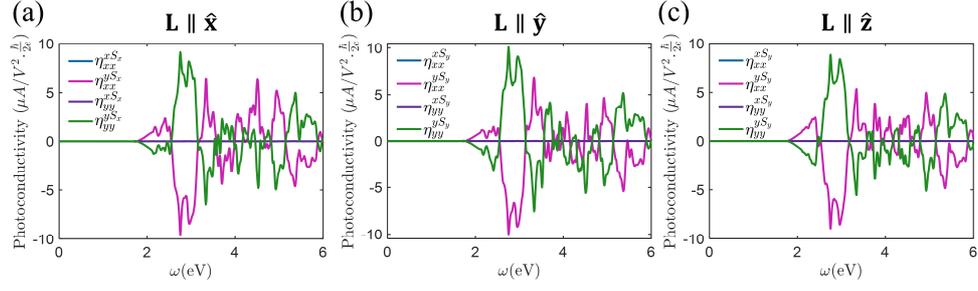

FIG. S4. Calculated spin current photoconductivity under $x$ and $y$-polarized LPL (a) $\mathbf{L} \parallel \hat{\mathbf{x}}$, (b) $\mathbf{L} \parallel \hat{\mathbf{y}}$, and (c) $\mathbf{L} \parallel \hat{\mathbf{z}}$.

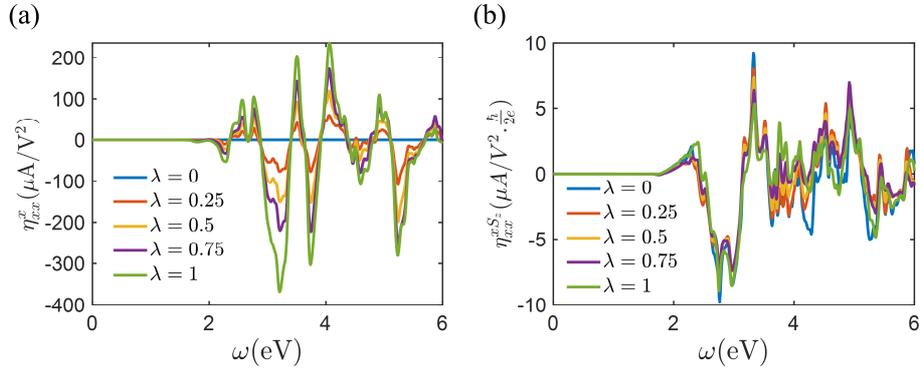

FIG. S5. (a) MIC photoconductivity $\eta_{xx}^{x}$ and (b) spin current photoconductivity $\eta_{xx}^{xS_x}$ under $\mathbf{L} \parallel \hat{\mathbf{z}}$ when SOC coefficient $\lambda$ increases from 0 (SOC totally turned off) to 1 (full SOC is included). One sees that the charge current almost linearly increases with SOC effect, while the spin current is not largely affected.

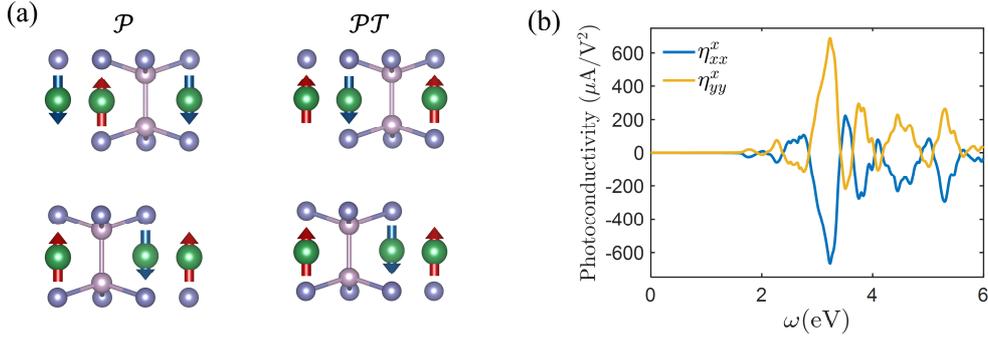

FIG. S6. (a) Bilayer MnPSe₃ structure in $\mathcal{P}$ and $\mathcal{PT}$ magnetic stacking patterns. They only differ in magnetic configurations, while the atomic coordinates are the same. (b) MIC of the $\mathbf{L} \parallel \hat{\mathbf{z}}$ in the $\mathcal{PT}$ magnetic pattern. The $\mathcal{P}$ magnetic pattern gives zero BPV photoconductivity.